\documentclass{article}

\usepackage{amssymb,latexsym,amsmath}

\usepackage[pdftex]{graphicx}

\usepackage{float}

\usepackage{hyperref, xcolor}

\hypersetup{backref, colorlinks=true, 
linkcolor=red, 
linkbordercolor = red,          
citebordercolor = blue}

\begin{document}

\pagestyle{plain}

\newtheorem{theorem}{Theorem}[section]

\newtheorem{proposition}[theorem]{Proposition}

\newtheorem{lema}[theorem]{Lemma}

\newtheorem{corollary}[theorem]{Corollary}

\newtheorem{definition}[theorem]{Definition}

\newtheorem{remark}[theorem]{Remark}

\newtheorem{exempl}{Example}[section]

\newenvironment{exemplu}{\begin{exempl}  \em}{\hfill $\square$

\end{exempl}}

\renewcommand{\contentsname}{ }

\title{GLC actors, artificial chemical connectomes, topological issues and knots}

\author{{\large Marius Buliga  and  Louis H. Kauffman}\\  \\
 \texttt{Marius.Buliga@gmail.com} ,  \texttt{kauffman@uic.edu}}

\date{ }

\maketitle

\begin{abstract}
Based on graphic lambda calculus, we propose a program for a new model of asynchronous distributed computing, inspired from Hewitt Actor Model, as well as  several investigation paths, concerning how one may graft lambda calculus and knot diagrammatics. 
\end{abstract}

\section{Introduction}
\label{proper}

Recent work on distributed computing takes inspiration from biology and chemistry. For example, the BIONETS collaboration \cite{bionets} propose biologically inspired "autonomic networks and services", based on fraglets \cite{tschudin1} and metabolic approaches \cite{tschudin2}. Decades earlier, Ban\^{a}tre, Le M\'{e}tayer et al \cite{banatre2} \cite{banatre1} introduced the chemical programming model of computation. Berry and Boudol \cite{bb} proposed the CHAM ("chemical abstract machine"), which  uses a chemical metaphor for modeling  asynchronous concurrent computations (in particular a concurrent lambda calculus).  Algorithmic Chemistry was introduced by Fontana and Buss \cite{fonbas1} \cite{fonbas2} \cite{fonbas3}.  

We borrow from this line of research the leading idea that lambda calculus is a kind of natural formalization of the bare bones of  chemistry, but we take it much further, starting from the following key observation. In a biological connectome,  neurons exchange electrical signals through synapses. At closer inspection, these signals are an effect (and a small part) of the chemical connectome (CC)  which runs in the background. The CC is  globally seen as a huge chemical reaction network (CRN) made of many  elementary, identical CRNs,  each one running on it's own and reacting to the chemical environment, in a kind of an asynchronous distributed computation. 

The key observation is that the global CRN is nothing but a god's eye view of this, instead the system functions exclusively by local interactions assembled into local metabolic cycles.  The signal transmission from a neuron to another is  an effect of  a swarm of localized  cycles of chemical reactions among molecules . This is in stark contradiction with the usual view which consists in  thinking in terms of signals passing through gates. It is instead the natural point of view of graphic lambda calculus and the chemical concrete machine.

Graphic lambda calculus (GLC) \cite{bgraph} \href{http://chorasimilarity.wordpress.com/graphic-lambda-calculus/}{(web tutorial)} is a graph rewrite system. Programs are certain trivalent graphs, and execution of programs means the application of graph rewrites, called "moves", on the respective graph. 

In the GLC formalism there is one global move, all the other moves are local (i.e. they involve a fixed, small number of nodes). There is  a variant of GLC, which uses only local moves,  called the Chemical concrete machine (chemlambda) \cite{buligachem} \href{http://chorasimilarity.wordpress.com/chemical-concrete-machine/}{(web tutorial)}. The moves of chemlambda act on graphs called "molecules" at certain "reaction sites", like chemical reactions involving molecules and enzymes (here  enzyme=move).

The execution of programs can be made into an asynchronous distributed computation, by looking at how real molecules react.

Real chemical interactions happen between  molecules which are close one to another. These proximity relations have to be a part of the computation model, somehow.  We need  to have a purely local mechanism for deciding  which parts of the graph are going to interact and how. We propose to transform molecules into actors and proximity relations into actors interactions.  Distributed computation is then seen as parallel, asynchronous application of reduction moves to a big graph which is split into smaller graphs, which interact as actors in an actor model  described in the section \ref{sactors}, which is inspired by  Hewitt \cite{hewitt1} \cite{hewitt2} \cite{actor}.

This opens the following possibility: the www is an artificial, human-made network and the GLC or the  chemical concrete machine are variants of an artificial, human-made, computing friendly chemistry. We may try then to use GLC actors  to endow the net with a chemistry, first, then exploit the chemistry to give the net a metabolism. This would lead to the construction  of Artificial Chemical Connectomes (ACC) which follow the artificial chemistry rules of GLC.

Let us enumerate the features of GLC and chemlambda which are significant: 
\begin{enumerate}
\item they allow for asynchronous distributed computation, not  seen as a Chemical Reaction Network (CRN), but instead one involving the graphs of GLC or the molecules of chemlambda, with chemical reactions driven by enzymes (i.e. moves) acting at reaction sites of these artificial molecules. Molecules are seen as actors and most of these chemical reactions are seen as actors interactions. 
\item they separate the essence of computation (here the application of moves) from the evaluation (here seen as a propagation of certain decorations of the graphs, according to certain local rules of decoration). Evaluation is not necessary for computation.
\item a part of the GLC graphs, or "molecules" from chemlambda, form a sector which represents untyped lambda calculus (without eta reduction).
\item  but there are other related sectors, like the one which represents the knot diagrams 
rewrite system, thus establishing a continuation of the research about knots and lambda calculus  \cite{kauffman2}, knot automata  \cite{kauffman3}, or about topological quantum computing \cite{chen}, \cite{kauffman1}. 
\item there is no variable (or terms) name or management of names. 
\end{enumerate}

The key merit of this model is a graphical reformulation of
the well-known lambda calculus, central to logic and to the design of
recursion in computer languages. By reformulating the lambda calculus in
terms of graphs, the operations for the calculus become essentially local
operations of graphical replacement. The graphs themselves
contain all the data that is usually formulated in terms of algebra. This means that
the global structure of the graph contains all the information that is
usually cut up into bits of algebra. The graph becomes a whole
system that instantiates the computational power of the calculus. This
instantiation is the key reason why this model can propose significant
designs in distributed computing. The graph as a whole can exist in a
widely distributed fashion, while the interactions that constitute its
computations are controlled by local nodal exchanges between actors. 

Furthermore, this property of redesigning the relationship of the local and the global is
not restricted just to lambda calculus networks. There are relationships
of the same kind that link this research with topology, topological
quantum field theory and quantum computation. 

Even more generally, the
movement between graphs and algebra is part of the larger picture of the
relationship of logical and mathematical formalisms with networks and
systems that was begun by Claude Shannon in his ground-breaking discovery
of the relationship of Boolean algebra and switching networks. We believe
that our graphical formulation of lambda calculus is on a par with these
discoveries of Shannon. We hope that the broad impact of this proposal will be a
world-wide change in the actual practice of distributed computing. Implemented successfully,
this proposal has a potential impact on a par with the internet itself.

\paragraph{Acknowledgments.} We were motivated to write this with the occasion of the preparation of a project  which  involves the authors, on the research side,  and a team from Proven Secure Solutions (PSS) on the side of a possible IT implementation. We thank  Stephen P. King from PSS, who made the first suggestion that   graphic lambda calculus might be used more easier on the net than, as one of the authors thought, in relation to real biochemistry, and to Jim Whitescarver from PSS for finding ways to make this type of computing real.

\section{GLC and chemlambda}

Graphic lambda calculus (GLC) \cite{bgraph} is a graph rewriting system.  GLC uses a set $GRAPH$ of oriented, locally planar trivalent graphs which are constructed from the following elementary nodes: (a) the $\lambda$ abstraction node, (b) the fan-out node, (c) the application node, (d) the dilation node, decorated with $\varepsilon \in \Gamma$, a commutative group of scales. 
\begin{figure}[H]
     \begin{center}
     \includegraphics[width=  80mm]{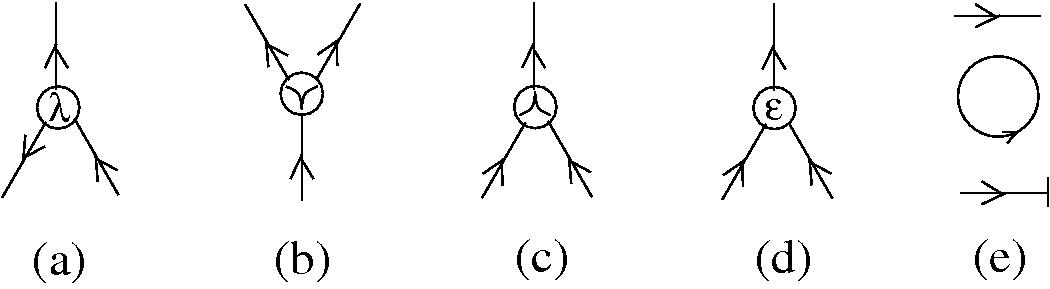}
     \caption{Basic pieces of GLC graphs}
     \label{4_nodes}
\end{center}
\end{figure}
To this nodes are added (e) arrows, loops and  a termination node with one incoming arrow and no output arrow. 

 The most important move (graph rewrite) is the graphic beta move.   It is the graphic version of beta reduction from lambda calculus. This is a local move, i.e. it affects only a local region of a graph. The graphic beta move is a purely oriented graph rewrite (move) version of the Wadsworth \cite{wads} or Lamping \cite{lamping} beta reduction move. 

\begin{figure}[H]
     \begin{center}
     \includegraphics[width=  80mm]{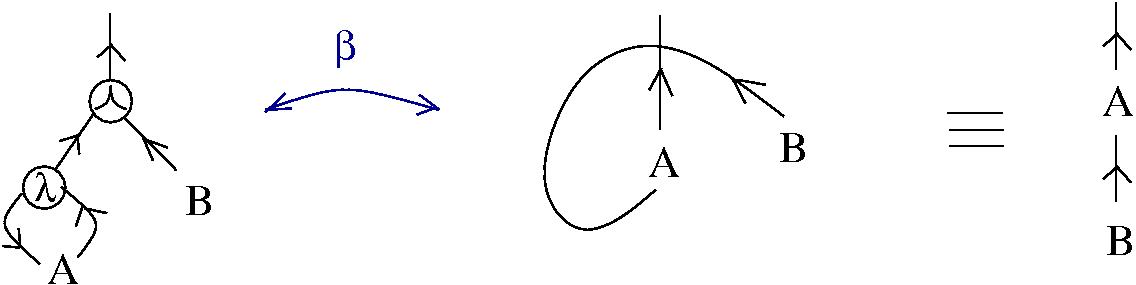}
     \caption{Standard application of the graphic beta move. From left to right, the graph of $\left(\lambda x.A\right)B$ becomes the graph of $A[x:=B]$}
     \label{only_beta}
\end{center}
\end{figure}

 There is no restriction though to apply this move only to lambda graphs (i.e. graphs which represent lambda calculus terms). Two examples are given in the figure \ref{all_moves_beta}: 

\begin{figure}[H]
     \begin{center}
     \includegraphics[width=  100mm]{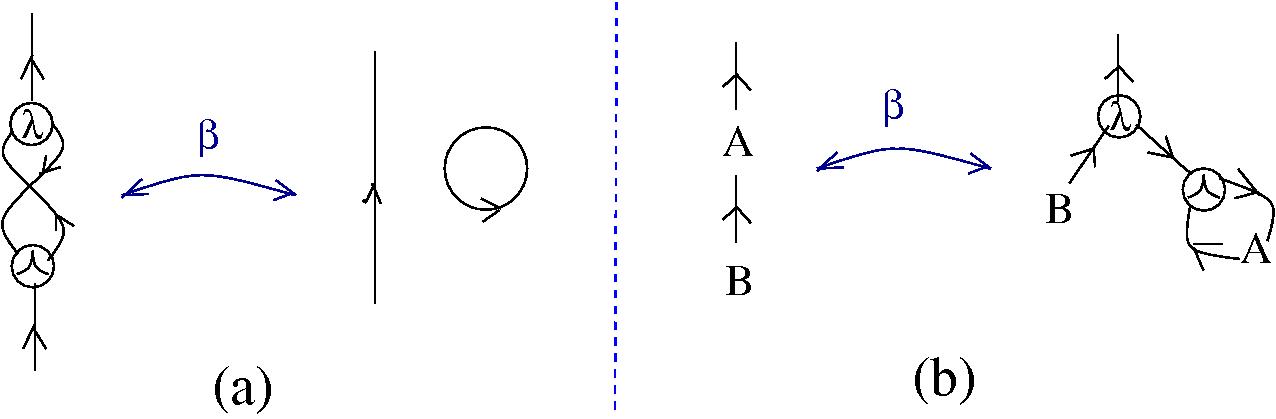}
     \caption{Non-standard applications of the graphic beta move: (a) loop with no nodes appears, (b) looks like the figure \ref{only_beta}, but the graph from the right is not one of a lambda term.}
     \label{all_moves_beta}
\end{center}
\end{figure}

\begin{figure}[H]
     \begin{center}
     \includegraphics[width=  80mm]{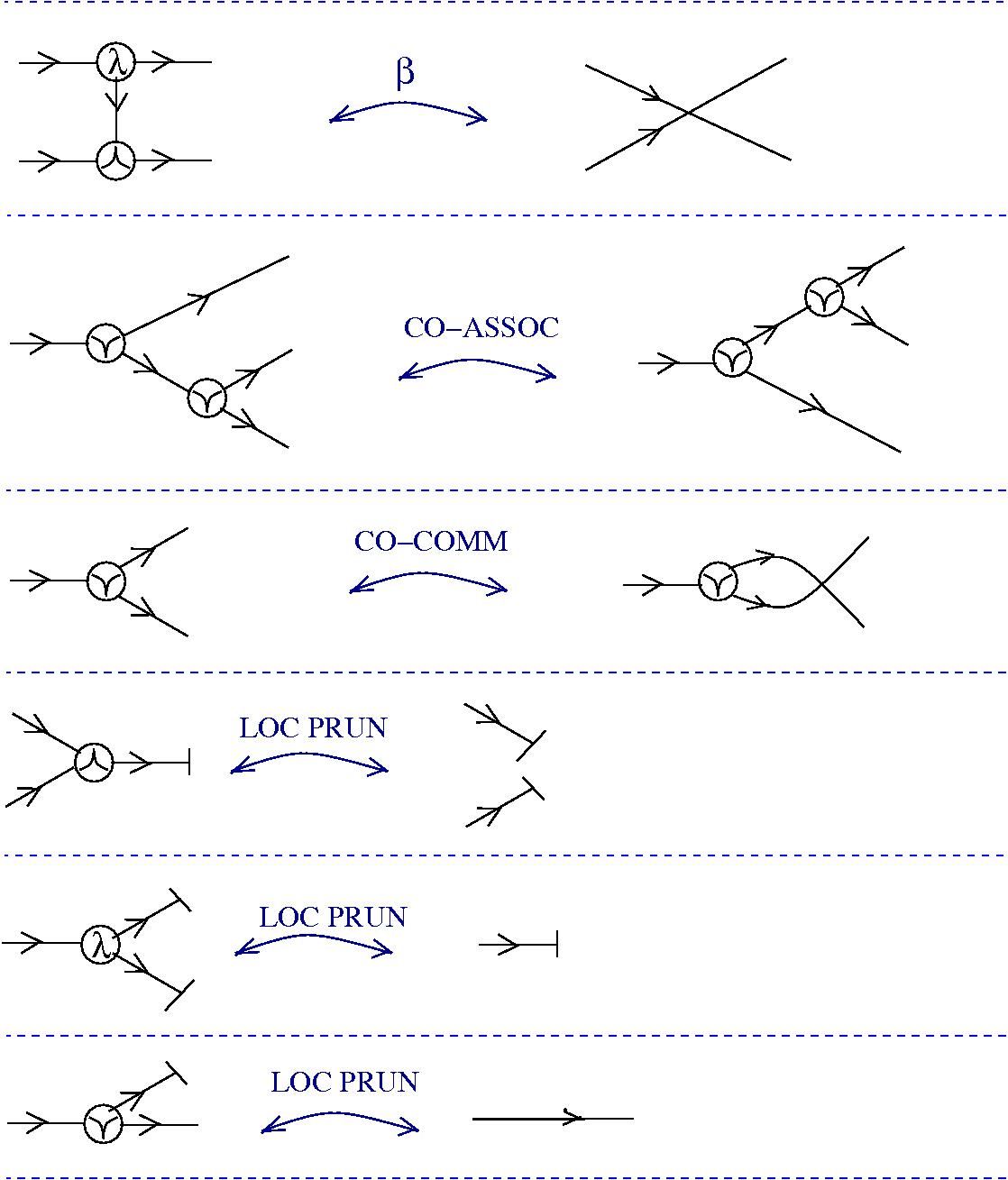}
     \caption{Local moves of GLC}
     \label{all_moves_glc}
\end{center}
\end{figure}

The moves of GLC come in two categories: local (with an upper bound on the number of nodes and arrows involved) and global (otherwise).  The list of all local moves of GLC which we shall use in this proposal  is given in the figure \ref{all_moves_glc}. (There are other moves, called R1a, R1b, R2 and ext2, which apply to dilation, termination and fan-out nodes, which will not be used in this note.)

\begin{figure}[H]
     \begin{center}
     \includegraphics[width=  90mm]{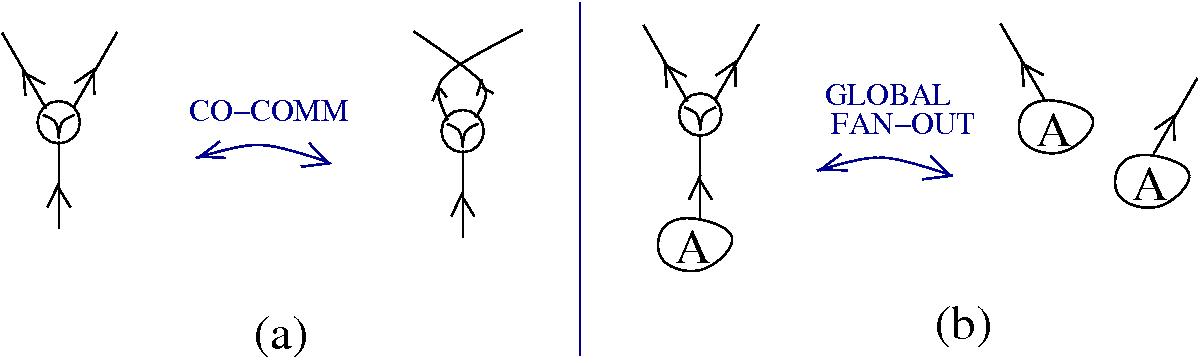}
     \caption{(a) the CO-COMM move is local, (b) the GLOBAL FAN-OUT move is global}
     \label{loc_vs_glob}
\end{center}
\end{figure}

To better understand the difference between local and global moves, look at the  figure \ref{loc_vs_glob}, which describes (a) the local CO-COMM move  and (b) the GLOBAL FAN-OUT move. 

Graphic lambda calculus has several interesting sectors (i.e. subsets of $GRAPH$ with particular choices of moves) which are equivalent with: (a) untyped lambda calculus, combinatory logic (b) knot and tangle diagrams with Reidemeister moves (c) finite difference calculus in spaces with dilations. 
\begin{figure}[H]
     \begin{center}
     \includegraphics[width=  90mm]{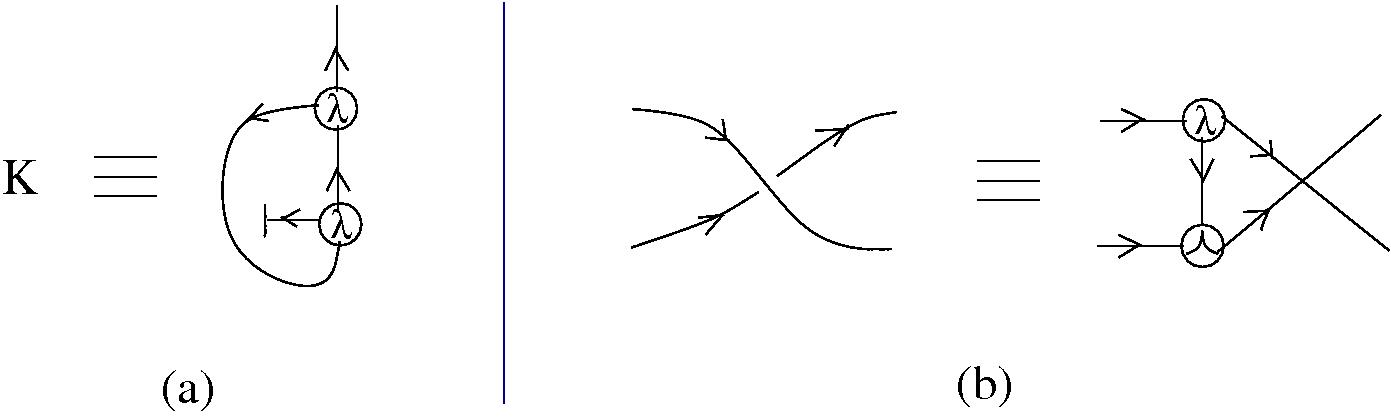}
     \caption{(a) the K combinator, (b) encoding of a crossing in GLC}
     \label{ex_macros}
\end{center}
\end{figure}

In the  figure \ref{ex_macros} are represented graphs from the first two  sectors: (a) the combinator $K$, (b) an oriented crossing.  The knot diagrams sector  allows GLC to interact with Kauffman Knot Logic  \cite{kauffman2} and Knot Automata \cite{kauffman3}, and  topological quantum computing in the sense of Kauffman and Lomonaco \cite{chen}, \cite{kauffman1}. This is discussed in section \ref{tglc}. 
 The motivation for constructing GLC was the need to have a visual representation of certain finite differential computations in spaces with dilations \cite{buligadil1}  \cite{buligainf}  \cite{buligaleng}\cite{buligabraided} \cite{buligairq}. This is possible in another sector of GLC, called the emergent algebra sector, which involves the dilation and fan-out nodes, along with CO-COMM, CO-ASSOC and the remaining moves R1a, R1b, R2 and ext2.

\bigbreak

The chemical concrete machine \cite{buligachem} (chemlambda) is a modification of graphic lambda calculus which uses only local moves on graphs which are called "molecules", following a chemical programming style. 
\begin{figure}[H]
     \begin{center}
     \includegraphics[width=  70mm]{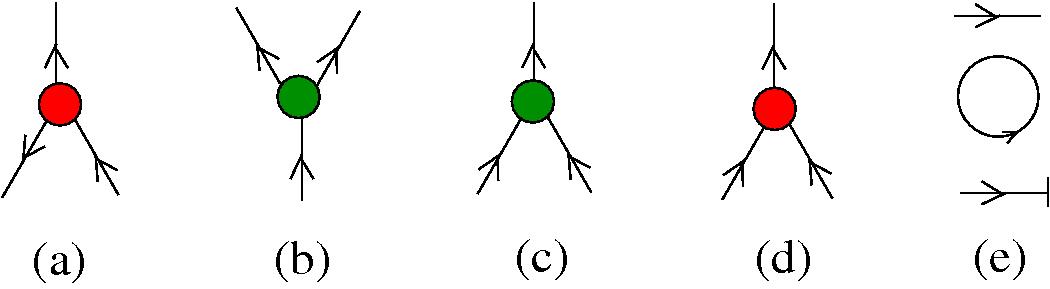}
     \caption{Basic pieces of chemlambda molecules}
     \label{4_chem}
\end{center}
\end{figure}

In chemlambda we admit also a set of nodes with unspecified valences, called "other molecules". These are the equivalent of "cores" from the section \ref{sactors}, paragraph 5. Interaction with cores, i.e. they can be used as interfaces with external constructs. All GLC local moves involving the $\lambda$ abstraction node, the application node and the fan-out node are the same in chemlambda. In the following figure we see the graphic beta move (at left), which is supplemented by a FAN-IN move (at right).   

\begin{figure}[H]
     \begin{center}
     \includegraphics[width=  100mm]{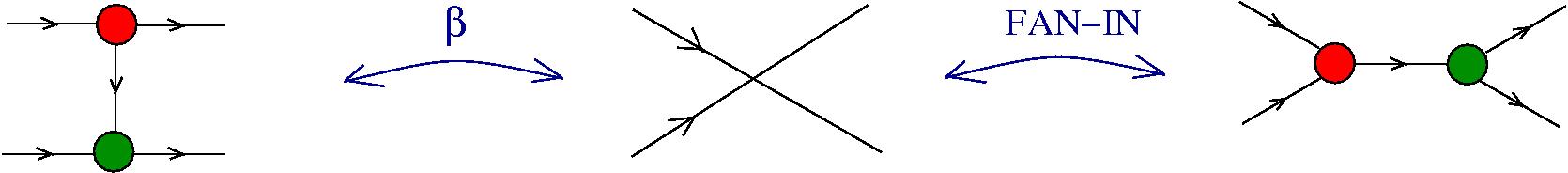}
     \caption{(left) the graphic beta move, (right) the FAN-IN move}
     \label{convention_2}
\end{center}
\end{figure}

The GLOBAL FAN-OUT move of GLC  is replaced by   the FAN-IN move and by two DIST moves, all local. 

\begin{figure}[H]
     \begin{center}
     \includegraphics[width=  70mm]{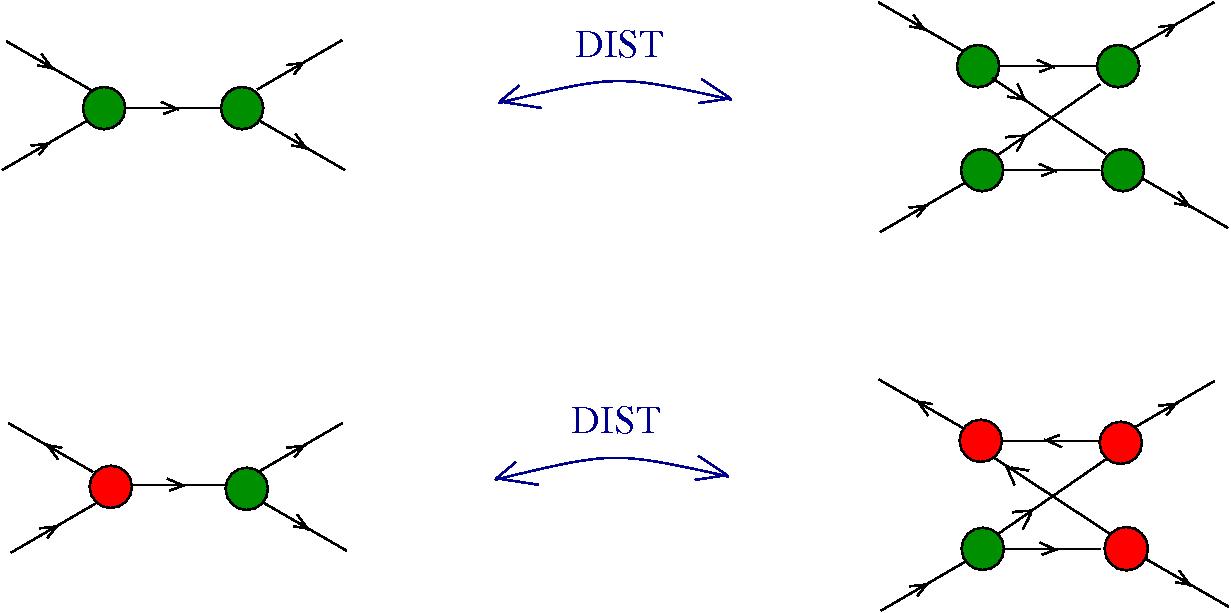}
     \caption{The local DIST moves}
     \label{convention_6}
\end{center}
\end{figure}

In the next figure we see an example of the replacement of GLOBAL FAN-OUT with a succession of moves in chemlambda. 
\begin{figure}[H]
     \begin{center}
     \includegraphics[width=  70mm]{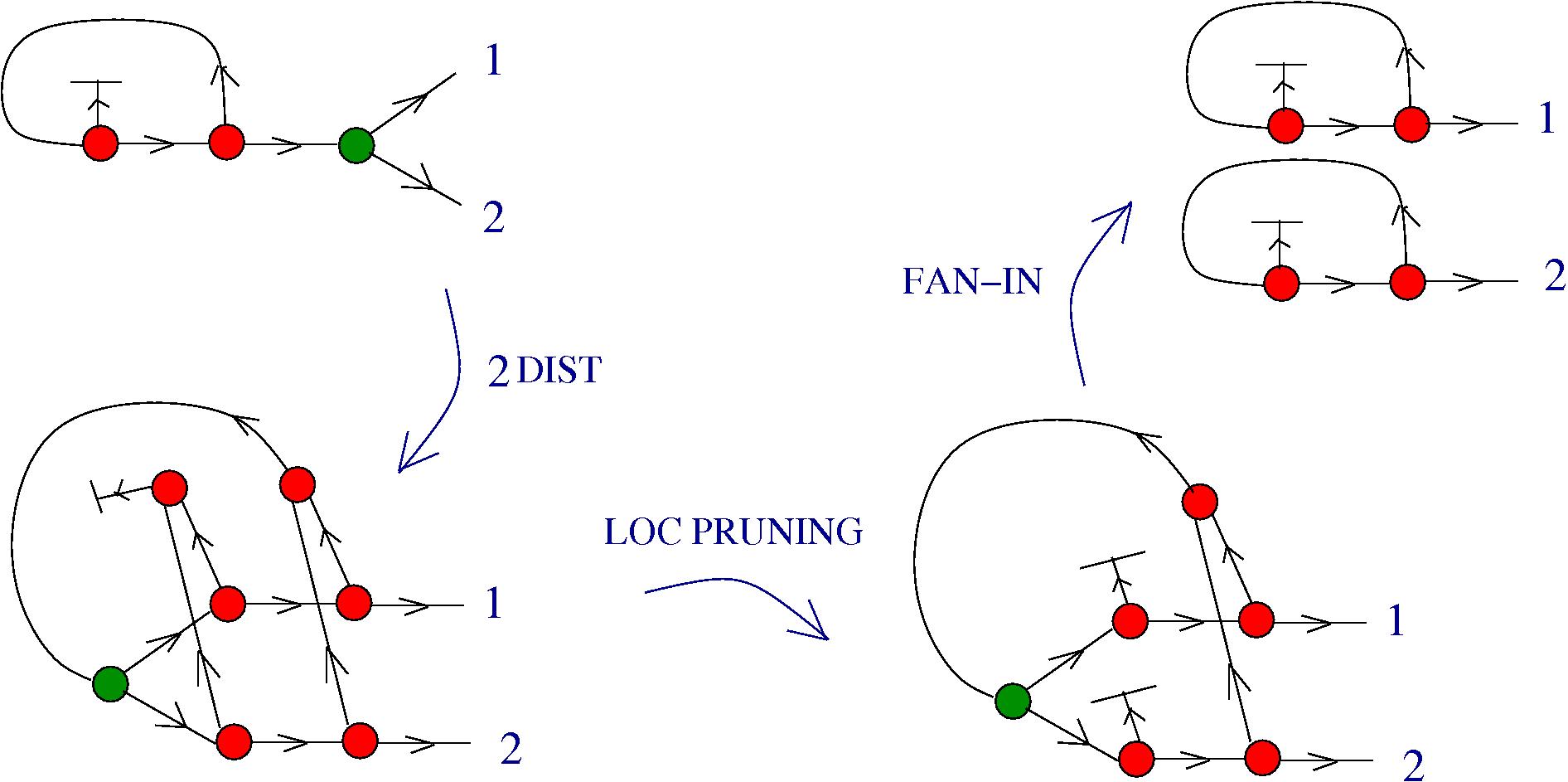}
     \caption{Example of use of local DIST moves for achieving a GLOBAL FAN-OUT move}
     \label{bckw_6_short}
\end{center}
\end{figure}

\section{GLC actors model}
\label{sactors}

Suppose we split a graph in $GRAPH$, or a molecule from chemlambda, into many parts, with the goal of reducing it in a distributed way. We take inspiration from Hewitt Actor Model \cite{hewitt2}, especially from  p. 14: " each beta reduction corresponds to an Actor receiving a message". 

Here is how we can achieve this. This computation model has two stages: preparation and computation.  In the first stage we define the actors (a procedure which is described here as a decoration of the GLC graph), then in the second stage, the actors interact according to 5 rules of behaviour, performing an asynchronous distributed computation. 

\begin{figure}[H]
     \begin{center}
     \includegraphics[width=  125mm]{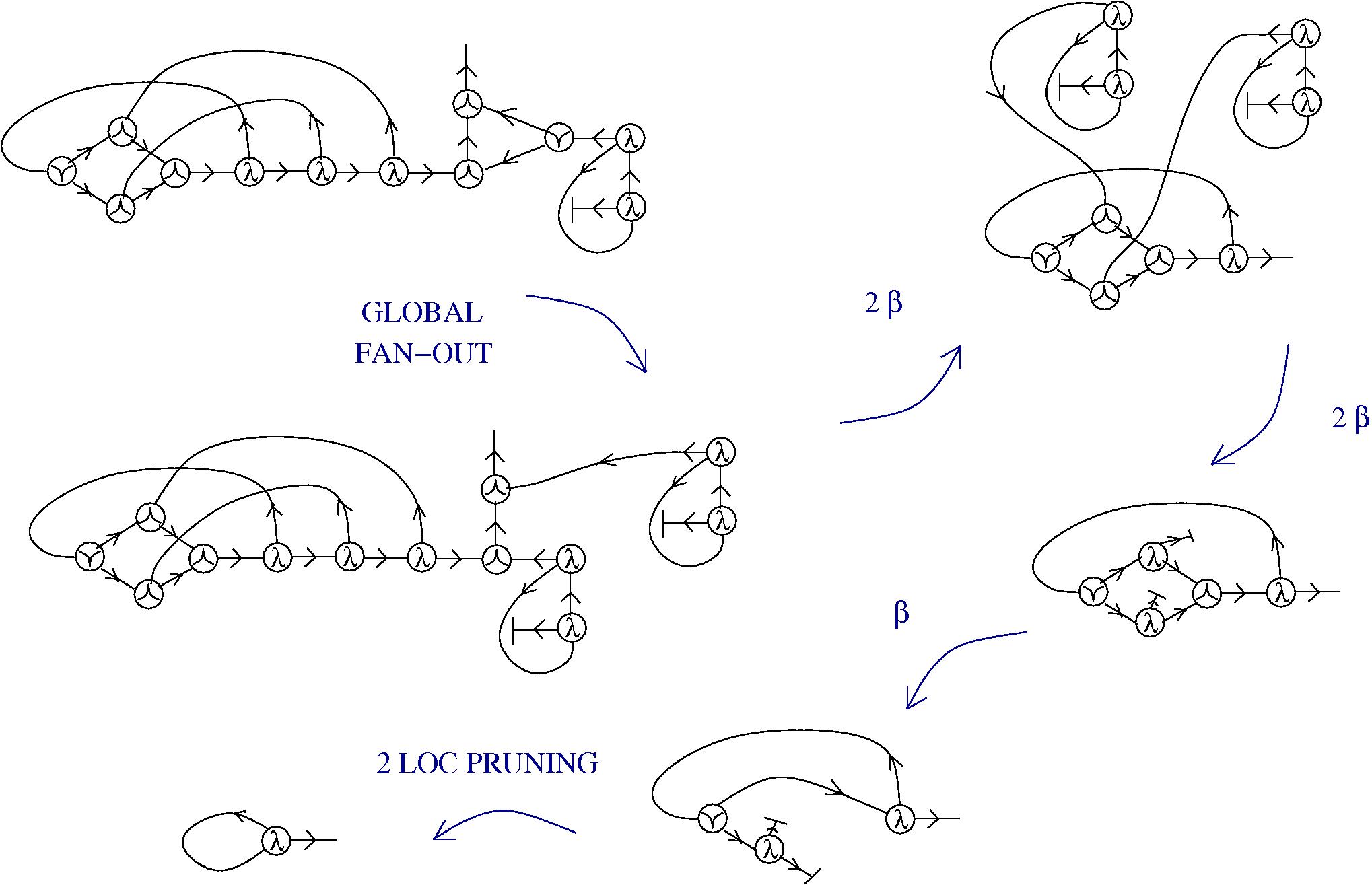}
     \caption{Reduction of the graph associated to $SKK$ to the graph associated to $I$}
     \label{skk_glc_crn_undec}
\end{center}
\end{figure}

We take as an example a graph which corresponds to the term $SKK$ in lambda calculus, and it's reduction to the combinator $I$, see figure \ref{skk_glc_crn_undec}.


\paragraph{Preparation.} We use a family of actors denoted by $a, b, c. ...$. Each actor $a$ has a name, or  address, denoted by $:a$. We shall decorate  the graph with  actor names. Then we decorate arrows with pairs of actors names. We  use the following notation for arrows decorations: $\displaystyle \langle :a \mid :b \rangle_{i}$, where the index $i$ is used when there are several arrows connecting the same pair of actors.  We suppose that $\displaystyle \langle :a \mid :b \rangle_{i} =  \langle :b \mid :a \rangle_{i}$.  Any arrow which joins two nodes decorated with different actor addresses is called a link between actors. 

Each actor will be in charge of the nodes and half-arrows which are decorated with the respective actor name. By definition, the actors diagram is the unoriented graph with actors as nodes and links between actors as arrows.

We may need to introduce, for practical reasons, supplementary entities, called "cores" which are not seen as graphs in GLC. (They may represent, in practice, interfaces with other computing entities, not necessarily implemented in GLC). The cores appear in this formalism as being nodes with in and out half-arrows, connected to the half-arrows of the graph in GLC.  The cores don't appear in the actors diagram because they are always wrapped into a GLC actor, called "mask".   For example, an actor $a$, has a name (address) $:a$, which may contain a mask (a graph in GLC) and a core $a^{*}$, such that all the half-arrows of the core are connected with some of the half-arrows of the mask; the remaining half-arrows of the mask are parts of the links of the actor $a$ with other actors.

\begin{figure}[H]
     \begin{center}
     \includegraphics[width=  80mm]{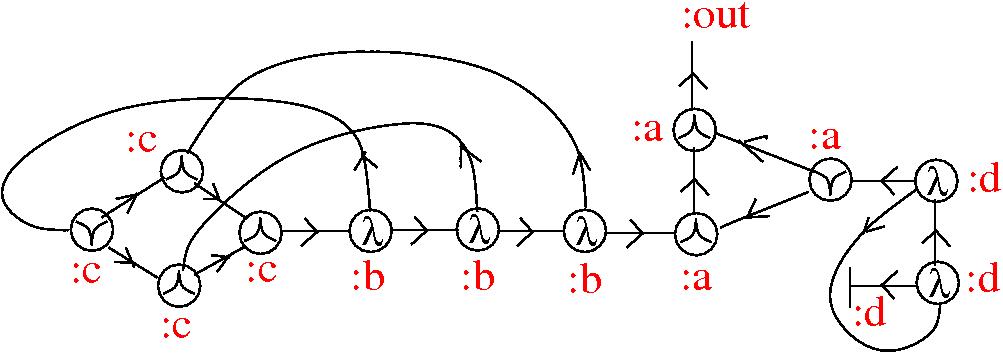}
     \caption{Decoration of the graph associated to $SKK$ by four actor names}
     \label{actor_prep_in}
\end{center}
\end{figure}

 For our example, in the figure \ref{actor_prep_in} we see the initial  graph, which appears in the upper left part of the figure \ref{skk_glc_crn_undec}, decorated with names of four actors. 

\begin{figure}[H]
     \begin{center}
     \includegraphics[width=  130mm]{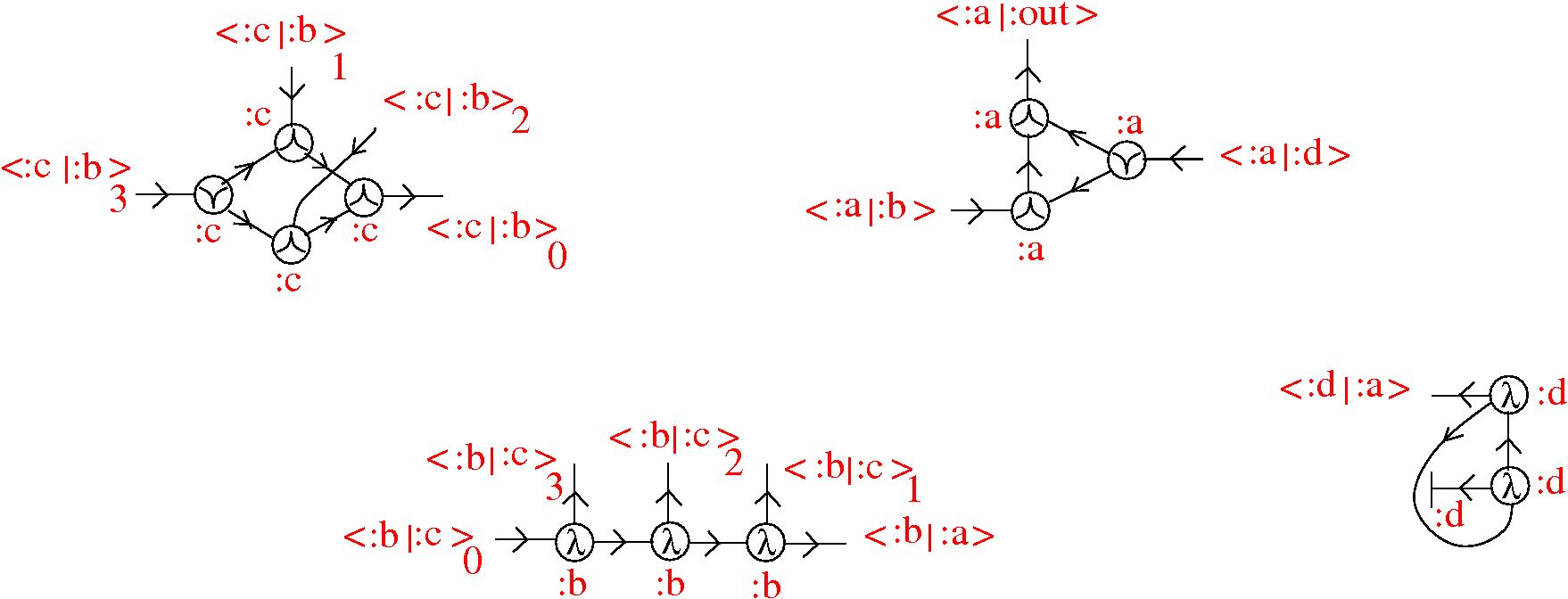}
     \caption{Definition of actors from the figure \ref{actor_prep_in}}
     \label{actor_prep_in_2}
\end{center}
\end{figure}

This decoration leads to the definition of the following four actors $a, b, c, d$, described in the figure \ref{actor_prep_in_2}. 

\begin{figure}[H]
     \begin{center}
     \includegraphics[width=  60mm]{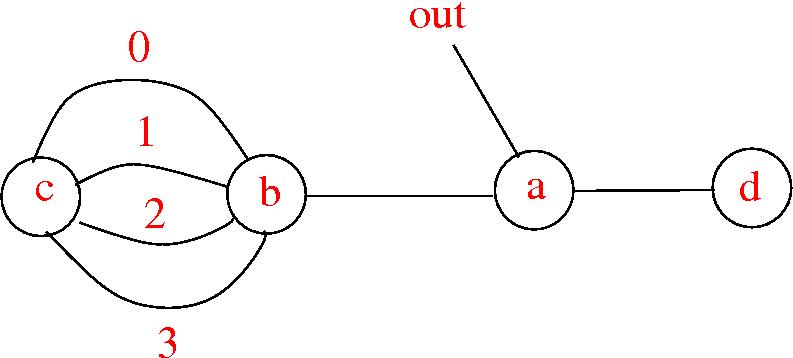}
     \caption{Initial actors diagram}
     \label{actor_prep_in_3}
\end{center}
\end{figure}

In the figure \ref{actor_prep_in_3} we see the initial actors diagram. 

\paragraph{Computation.} Starting from the initial actors, the computation consists into applying moves to the graph made by the union of actors (and possibly the cores), according to the following rules which describe what one actor can do. The interactions between actors or inside actors  are described by the way  decorations change when we apply a move in GLC. The possible actors behaviours are listed further. 

\smallbreak

\paragraph{1. Moves as interactions between actors.}  An actor tries to apply  reduction moves which involve a link with another actor. In the next figure we see an application of the graphic beta move between actors $a$ and $b$. The move affects the connectivity of  the actors $c$, $d$, $e$, $f$. 
\begin{figure}[H]
     \begin{center}
     \includegraphics[width=  90mm]{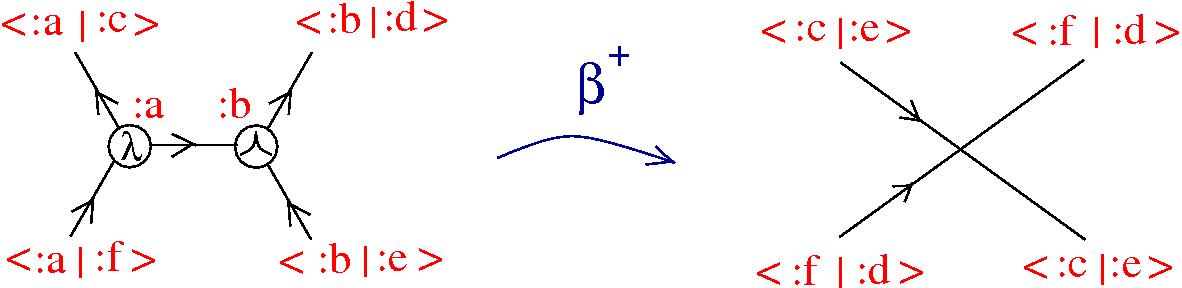}
     \caption{The graphic beta move as  interaction between the actors $a$ and $b$}
     \label{inter_actor_1}
\end{center}
\end{figure}

In general, there is no condition on the actors $a, b, ..., f$ from figure \ref{inter_actor_1} to be different. If the actors $a$ and $b$ are the same then we may speak about self-interactions. If chemlambda is used instead of GLC then we might consider the FAN-IN move (right side of the figure \ref{convention_6}) as an interaction between actors too. 
It is possible that, after an interaction as the one from figure \ref{inter_actor_1}, we obtain a loop, like in  the figure \ref{all_moves_beta} (b). The possibility of having actors with no nodes, only arrows, is left open. 

 There are multiple possibilities for the concrete mechanism of interaction, see section \ref{disc} for a discussion about this. 

\smallbreak

\begin{figure}[H]
     \begin{center}
     \includegraphics[width=  80mm]{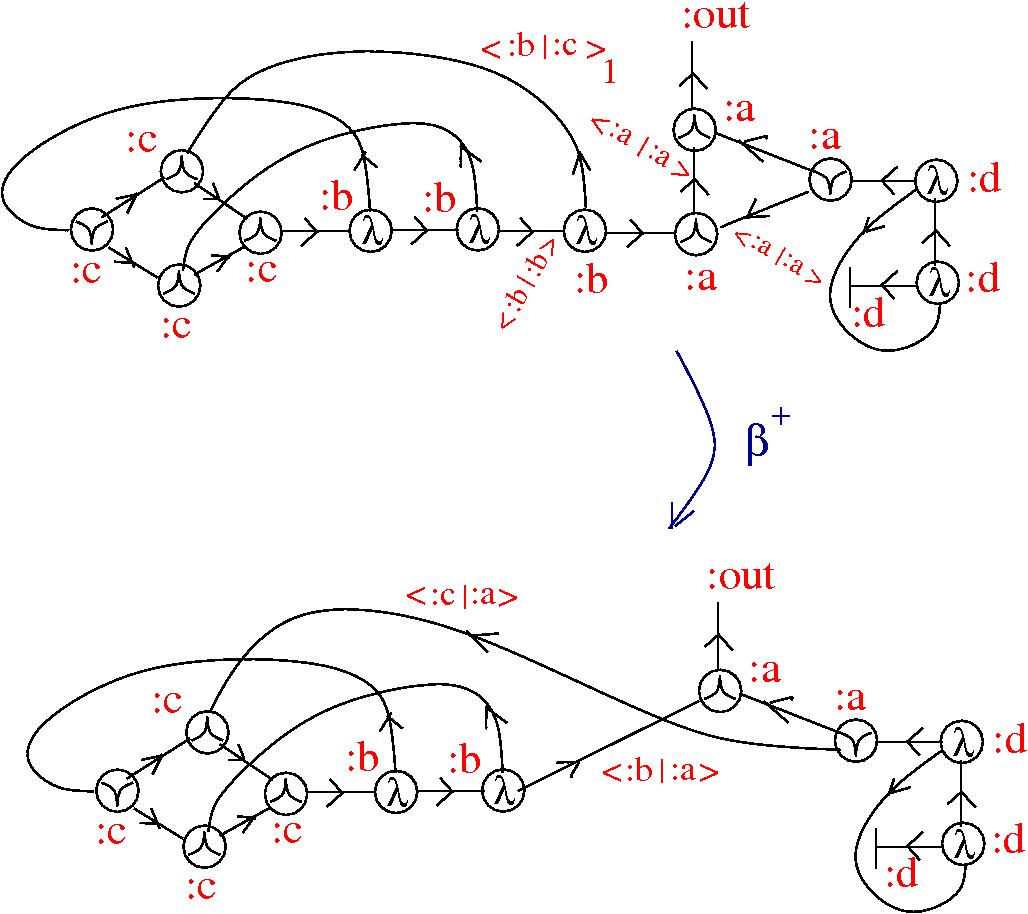}
     \caption{An example of interaction between two actors by a graphic beta move}
     \label{actor_prep_beta}
\end{center}
\end{figure}

In the case of the  example from  the figure \ref{actor_prep_in}, the actors $a$ and $b$ interact as in the  figure \ref{actor_prep_beta}. 

\begin{figure}[H]
     \begin{center}
     \includegraphics[width=  60mm]{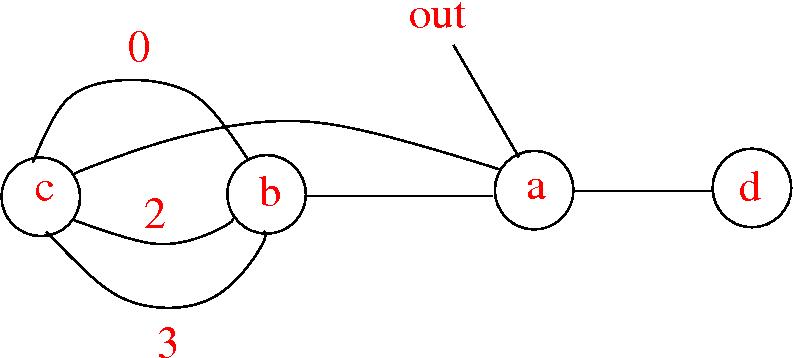}
     \caption{After the interaction from the figure \ref{actor_prep_beta}, the actors diagram from the figure \ref{actor_prep_in_3} changes into this one}
     \label{actor_mod_beta}
\end{center}
\end{figure}

After this interaction all the actors changed. Indeed, each of the actors $a$ and $b$ loose one node, but also the decorations of some arrows change; the actors $c$ and $d$ don't loose nodes, but the decorations of some of their arrows change. As an effect, the connectivity of the actors diagram from the figure \ref{actor_prep_in_3} changes after this interaction into the graph from the figure \ref{actor_mod_beta}. 

\smallbreak

\paragraph{2. Name change.} Nodes of an actor, like a fan-out node, or a termination node can change the name into one of the actors names which has a link in common with it. 
\begin{figure}[H]
     \begin{center}
     \includegraphics[width=  90mm]{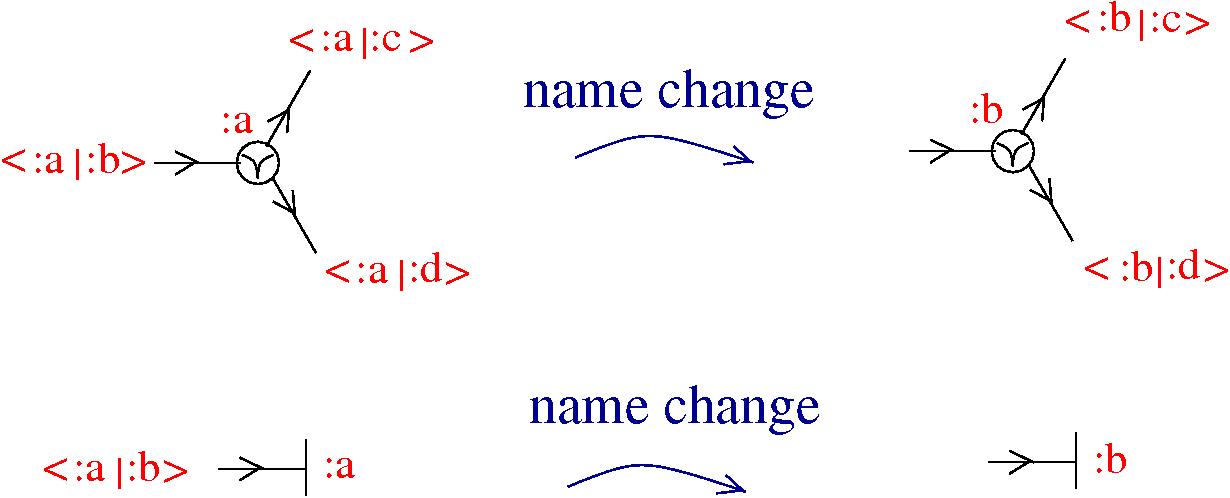}
     \caption{(up) Name change of a fan-out node, (down) Name change of a termination node }
     \label{inter_actor_2}
\end{center}
\end{figure}
We may interpret the upper part of the figure \ref{inter_actor_2} as $a$ sending to $b$ an order to produce a copy of $b$.   There is one half-arrow at the right upper part of the figure with missing decoration. That decoration has the form $\langle :b \mid ... \rangle$, with the missing name actor depending on the connectivity of the fan-out node along that arrow. 

\begin{figure}[H]
     \begin{center}
     \includegraphics[width=  90mm]{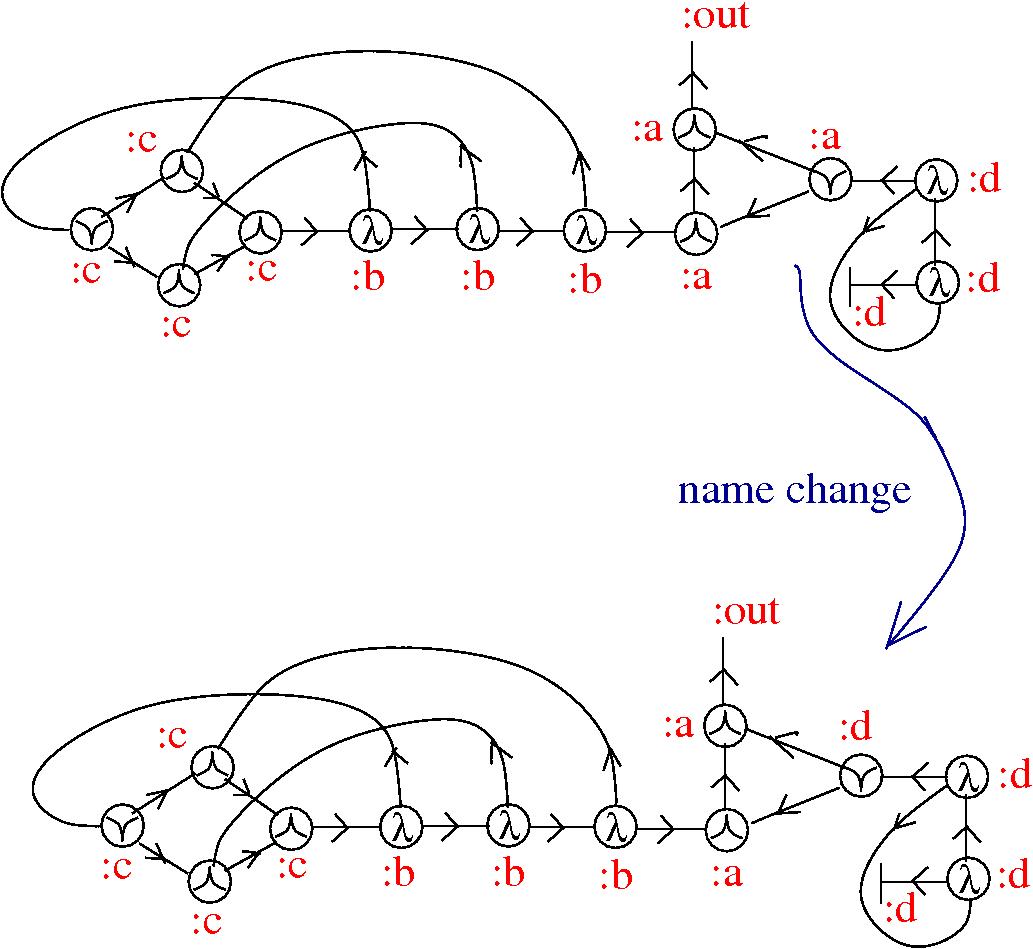}
     \caption{The actor $a$ sends to the actor $d$ and order to produce a copy of itself}
     \label{actor_skk_name_change}
\end{center}
\end{figure}

In our example, starting from the preparation from the figure \ref{actor_prep_in}, instead of interacting with $b$ (figure \ref{actor_prep_beta}) the actor $a$ might interact with the actor $d$ by a name change. This is described in the figure \ref{actor_skk_name_change}. 
After this name change move, the initial actors diagram from the figure \ref{actor_prep_in_3} becomes the following one, and not the one from the figure \ref{actor_mod_beta}: 
\begin{figure}[H]
     \begin{center}
     \includegraphics[width=  55mm]{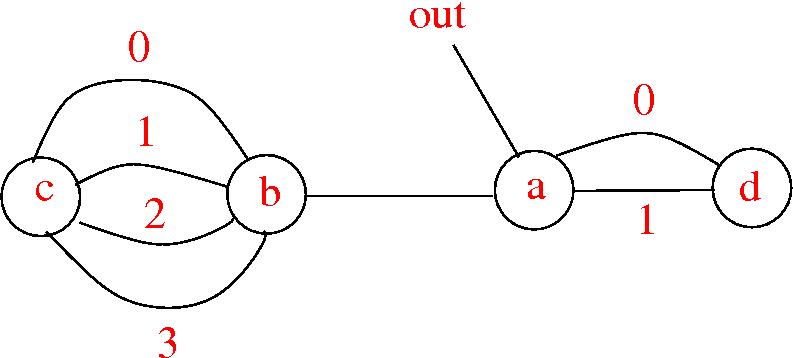}
     \caption{After the interaction from the figure \ref{actor_skk_name_change}, the actors diagram from the figure \ref{actor_prep_in_3} changes into this one, instead of the one from figure \ref{actor_mod_beta}}
     \label{actor_name_dia}
\end{center}
\end{figure}

The lower part of the figure \ref{inter_actor_2} can be interpreted as $a$ sending to $b$ a pruning command. The same comment, as previously, can be applied to the fact that there is a missing decoration of a half-arrow in the lower right part of the figure. 

We encounter this in our example of computation from the figure \ref{skk_glc_crn_undec}. At some point during the computation we need to apply a first local pruning move (lower left part of the figure  \ref{skk_glc_crn_undec}). Before doing the local pruning, we need first to do a name change, like in the following figure (the actors names are the ones which appear from the initial preparation stage from the figure \ref{actor_prep_in}; notice the appearance of a new actor $e$, coming from the creation of new actors explained at point 4 later). 
\begin{figure}[H]
     \begin{center}
     \includegraphics[width=  120mm]{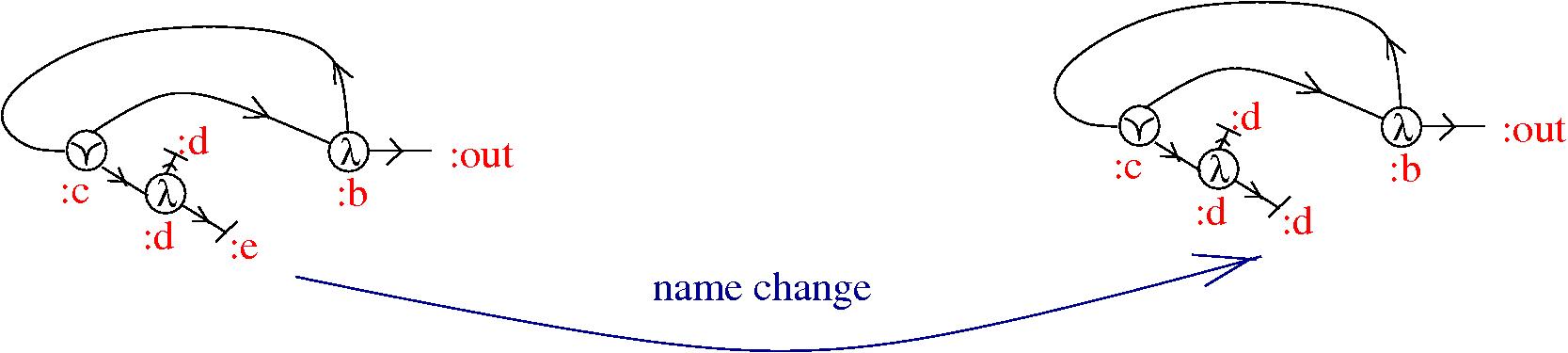}
     \caption{The actor $e$ sends to the actor $d$ and order to produce a pruning command}
     \label{skk_glc_crn}
\end{center}
\end{figure}

\smallbreak

\paragraph{3. Internal moves.} An actor can perform internally some moves. For example, upon receiving a fan-out node by a name change interaction,  the actor may produce such a copy by a GLOBAL FAN-OUT move inside the actor $b$, or, in case actors are implemented with chemlambda, there could be used DIST  moves instead, with the same effect. In our example, the actor $d$, upon receiving the fan-out node from $a$, can proceed  like in the figure \ref{bckw_6_short}, where the graph of the $K$ combinator is multiplied into two copies. Likewise, if the actor receives a pruning command then it can start to apply internally LOCAL PRUNING moves. 

\smallbreak

\paragraph{4. New actors.} An actor $a$ may have   its nodes partitioned  into two disjoint sets, with no arrow connecting nodes in different sets of the partition. Such a partition may appear after the actor performed internally a GLOBAL FAN-OUT, or after a name change interaction. The actor can then  create two new actors, one called $:a$ and the other one with a different name. In our example, the actor $d$, after doing internally a GLOBAL FAN-OUT move, splits into two actors, the first called $d$ and the other called $e$, both actors containing a copy of the graph of the combinator $K$.

\smallbreak

\paragraph{5. Interactions with cores.}  An actor with a core, if no other action is available, can "express" a part of the core, i.e. it can transform a part of the core into a part of its mask, or conversely, it can move a part of its mask into the core, according to the particular rules of the core (interface). 

\begin{figure}[H]
     \begin{center}
     \includegraphics[width=  120mm]{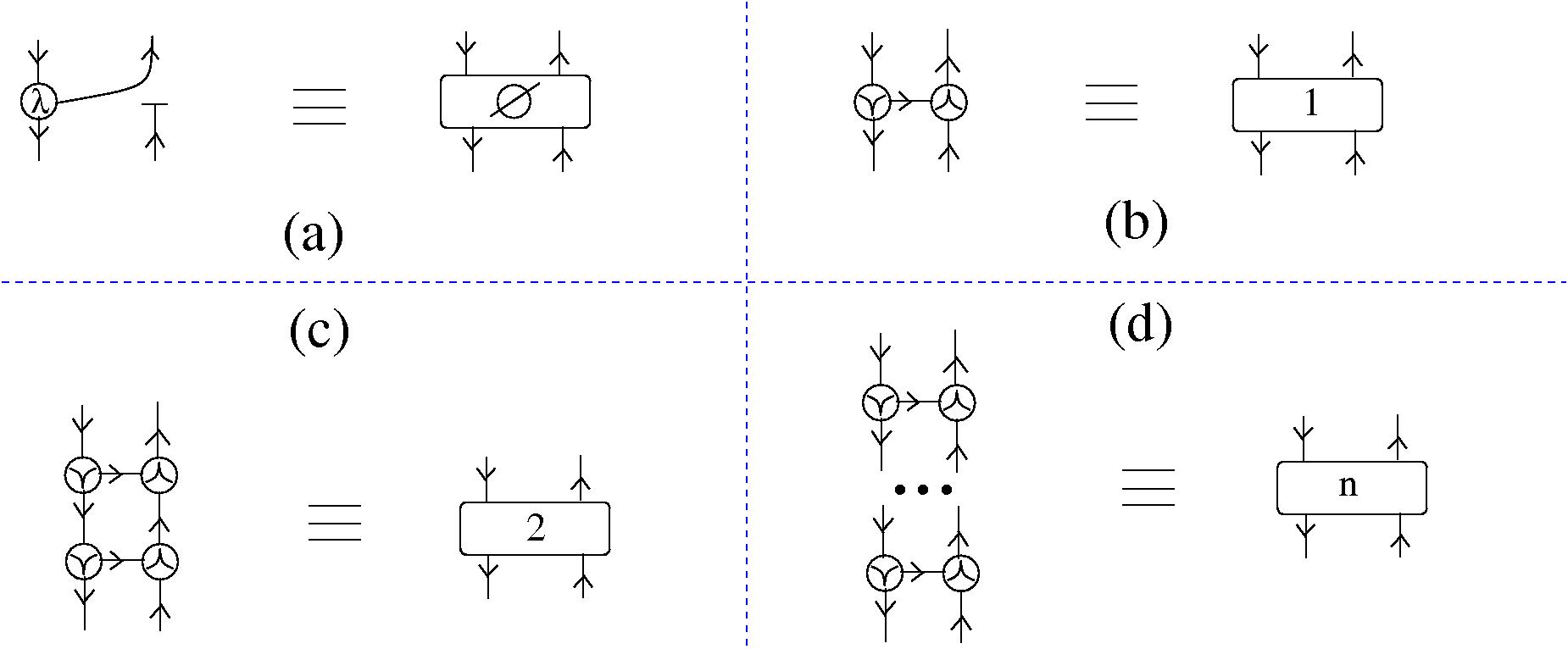}
     \caption{Stacks of units: (a) empty, (b) one unit, (c) two units, (d) $n$ units}
     \label{core_1}
\end{center}
\end{figure}

\begin{figure}[H]
     \begin{center}
     \includegraphics[width=  120mm]{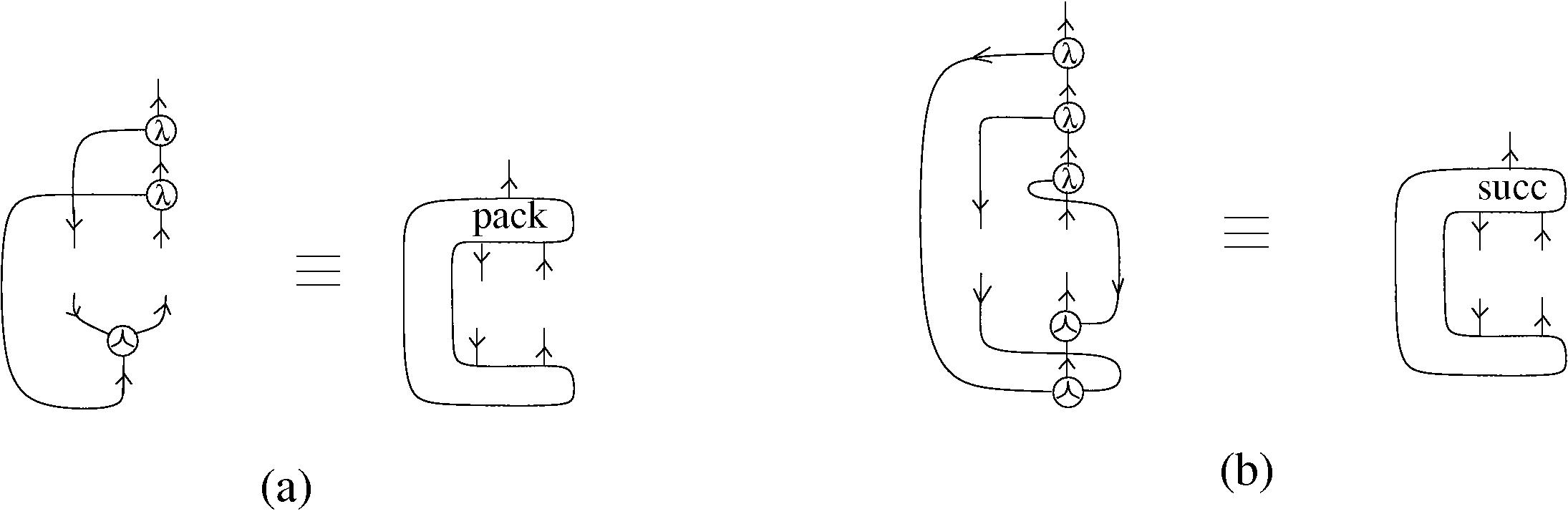}
     \caption{(a) the pack mask, (b) the successor mask}
     \label{core_2}
\end{center}
\end{figure}

\begin{figure}[H]
     \begin{center}
     \includegraphics[width=  100mm]{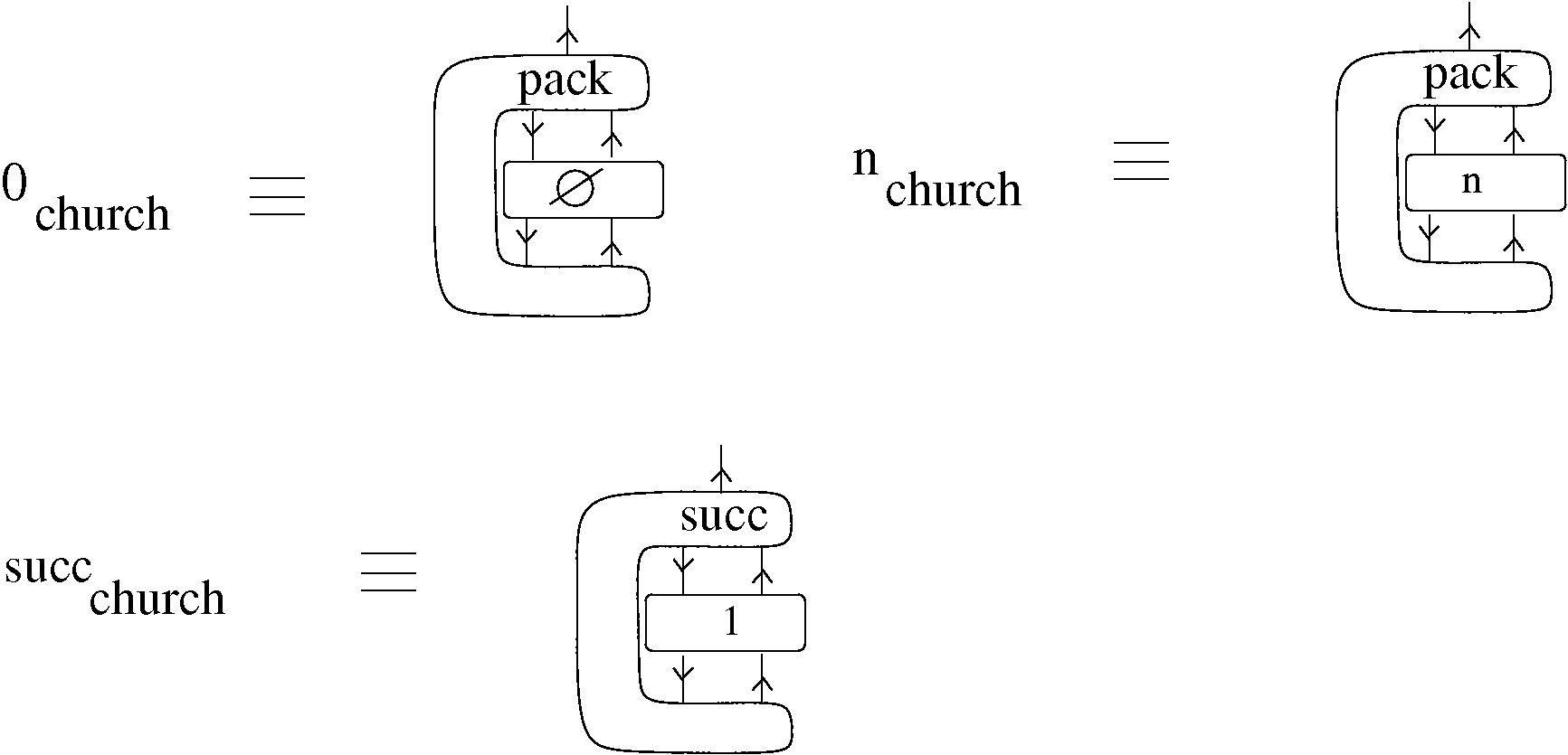}
     \caption{Church numerals and  the successor, built from masks and cores}
     \label{core_3}
\end{center}
\end{figure}

As an example, let us give the following description in GLC of the Church numerals: in the figure \ref{core_1} are introduced some stacks of "units", and in the figure \ref{core_2} we see the "pack" and "successor" masks. 
Then, in the figure \ref{core_3}, we see the graphs in GLC which correspond to the lambda terms for the Church numerals and  the successor. 
The stack of units is seen as a core of the actor which implements a Church numeral, having a pack mask. The stack may be e a counter, implemented possibly in another formalism than GLC.  (We can say the same about the successor, which can be seen as a pair mask-core; moreover, if the core corresponds to a counter which has a value different than one, then the successor still makes sense, but will behave as something which increments numerals by the value which is in the respective counter, or core.) 

\begin{figure}[H]
     \begin{center}
     \includegraphics[width=  55mm]{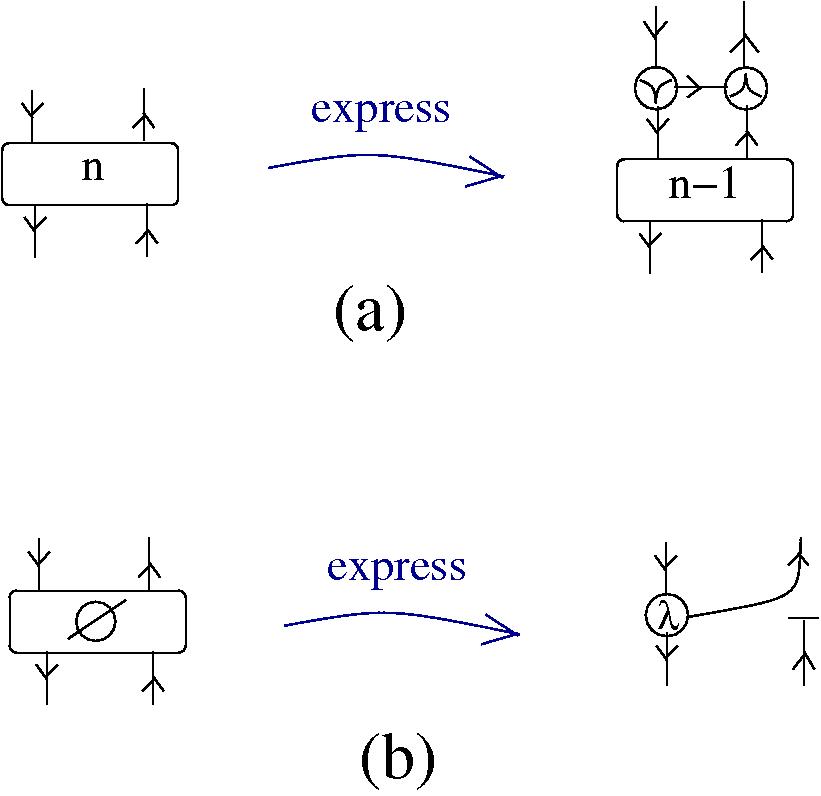}
     \caption{(a) if the value of the counter $n>0$ then the core expresses a unit and the counter decreases by one, (b) if the value of the counter $n =0$ then the counter disappears and it expresses an empty stack of units}
     \label{core_4}
\end{center}
\end{figure}

In our case, it makes sense to define two possible interactions with cores, like in the figure \ref{core_4}.

The interaction with cores is therefore designed from the knowledge of the core structure. Indeed, if the core represents data in some format, then we know that we can express the respective data format in lambda calculus, hence in GLC. From this point, the interaction with the respective core simply means a translation from the data format used by the core into GLC. 

\section{Discussion of this model}
\label{disc}

This is a model of distributed, asynchronous computing. It has a new feature which becomes obvious by looking at the messages exchanged by the actors. 

The preparation stage of the computation leads to the creation of the GLC actors. 
Once the preparation stage ended, the actors are left by themselves to interact. Let's examine their exchanges. 

\paragraph{ Interaction by the graphic beta move.} There are potentially 6 different actors implied in the interaction by the graphic beta move, see figure \ref{inter_actor_1}. As mentioned, there are several ways to perform this interaction. Any of them will have a comparable load of message passing. Let's look at one example. 

Suppose that the actor $a$, which has the $\lambda$ node, asks the actor named $:b$ what is the node connected with that node. He must send to $b$ something like the name $:a$ and some bits, maybe about the node $\lambda$. There are 3 bits needed to identify the node lambda, once the actor $b$ already knows the orientation of the arrow which links it with $a$. Two bits are needed to specify the orientation of the other arrows of the $\lambda$ node and one bit to tell if it a $\lambda$ node or a fan-out node (because both have the same pattern of arrows). 

This message is practically a packet of the form $(:a, BBB, :b)$ where $BBB$ are the three bits. The message content is very small. 

If the actor $b$ finds out that his application node forms a good configuration with the $\lambda$ node of the actor $a$, he may send the command to $:a$ to signal this (1 bit needed, sandwiched between the names $:b$ and $:a$) and then the actor $b$ proceeds by erasing it's application node and then by sending to $:d$ the label $\langle:f \mid :b\rangle$ and to $:e$ the label  $\langle :c \mid :b\rangle$. 

Upon receiving these messages, the actors $d$ and $e$ update their labels by concatenation: the label $\langle :b \mid :d\rangle$ becomes 
$$\langle:f \mid :b\rangle \langle :b \mid :d\rangle \rightarrow \langle :f \mid :d \rangle$$
and the same for the label of the half-arrow of $e$. 

The same is done by the actors $a$, $c$, $f$. 

Upon messages for confirmation that the actors $c$, $d$ $e$, $f$ updated their labels, the actors $a$ and $b$ may forget they were involved in the links between $c$ and $e$ and respectively between $f$ and $d$. 

\paragraph{Name changes.} These involve only interaction between two actors, figure \ref{inter_actor_2} and are thus much more simple than the previous one. Only the information about the node which is exchanged are in the messages, with a confirmation of the exchange performed. 

\paragraph{New actors, interaction with cores.} This two interactions are more complex, because in the case of producing new actors, as well as in the case of interaction with cores, what is happening is that possibly many nodes and arrows change names, or are produced. The interaction with cores involves one actor and it's core, therefore there are exchanges only between those. Creation of new actors supposes the creation of a new actor (machine) and then a name change interaction between the old actor and the new actor.

\paragraph{Internal moves.} They don't involve interactions between actors. 

\bigbreak

The main new feature of this model is that there are no complex messages exchanges between actors (with the exception of new actors creation and interaction with cores). Moreover, the messages are not circulating much through the actors network. 
Finally, even when considering internal moves, this is not a model based on signals circulating through wires, which are processed by gates (with the exception of interaction with cores).

\section{Topological Issues and Knots}
\label{tglc}

\begin{figure}[H]
     \begin{center}
     \includegraphics[width=8cm]{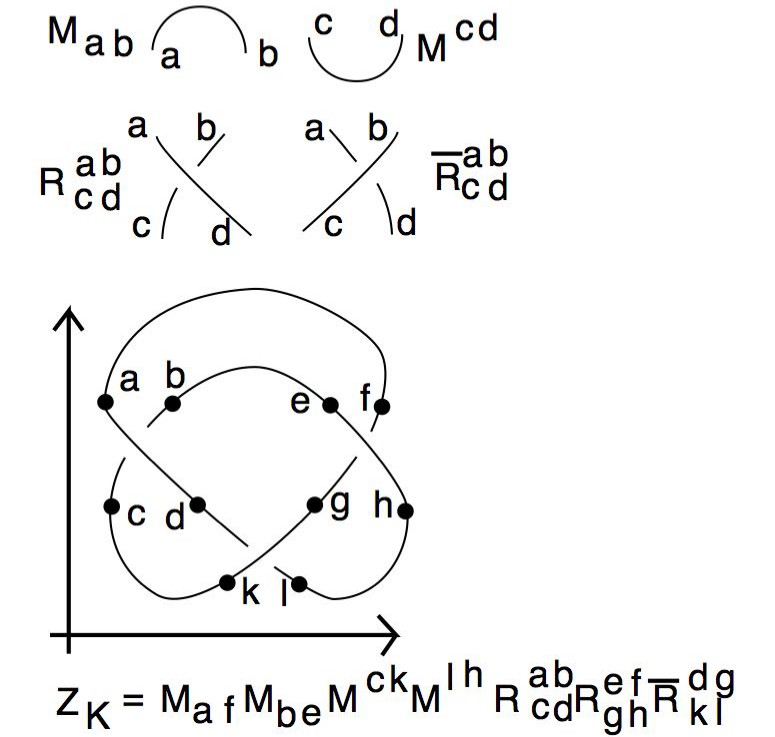}
     \caption{Knot Diagrams and Abstract Tensors}
     \label{abtens}
\end{center}
\end{figure}

The fact that alpha reduction is not needed in the GLC 
 due to the absence of variables and the presence of direct
connections that effect interactions is part of a link of this formalism
with the formalisms of topological quantum field theories at the knot
theoretic level. This sounds like a mouthful, but it is actually very
simple. The graphical issue is the same. Lets talk about knot and link
diagrams. In writing a knot diagram so that it can 'turn into' a quantum
link invariant one divides the diagram up into pieces, each of which is
in the form of a blackbox with (say) two or four lines emanating from it.
There is a black box for each type of crossing and special boxes for
maxima and minima. See Figure~\ref{abtens}. 
Once such a decomposition has been made, one can assign
abstract variables to each of the ends of these boxes. Boxes that are
connected to one another have lines that receive the same abstract
variables. We can then translate the knot diagram into an abstract tensor
expression with double appearances of indices connoting lines that are
connected to one another. The abstract tensor expression is the analogue
of the algebraic lambda calculus expression. One must take great care in the
substitution of dummy indices. Indices that label different lines in the
diagram must remain different. This leads to a non-local calculus of
substitutions in the abstract tensor algebra. So far we are only making an
algebraic image of the knot diagram. To get computable invariants we go
farther and map the abstract tensors to actual matrices with finite
numbers of index possibilities and tied indices are summed over just as in
ordinary tensor calculus. Thus the relationship of the logic of
substitution and the logic of graphic and topological connection is very
close in these two subjects.
\bigbreak




The issue of non-local substitution in systems is a form of paradox
protection that we have elsewhere called the {\it Flagg Resolution.} If it
should be the case that $J$ is an entity in a logical algebra such that
$\sim J = J$ ($\sim$  denotes a negation operator of order two). Then one regards $J$ as
a ``hot potato" and agrees that the substitution of $\sim J$ for $J$, if effected, in
an expression, must be so effected for every occurrence of $J$ in the
expression. This is a non-local rule that is directly enforced in a
graphical version of the algebra by having exactly one entity $J$ and
connecting edges from the $J$-node to all of the former appearances of $J$ in
the algebra. This means that problems of typing in lambda calculus will
also take on a different aspect once the calculus is graphical. We shall
examine this in depth in our research. We expect that the questions of
logical type will interlace with the recent work of Voevodvsky \cite{Voed} on
Homotopy Type Theory for mathematical foundations.
\bigbreak

One  difference between knot theoretic considerations and lambda
calculus considerations is in the fact that we do not usually think of a
knot diagram as a computing element that undergoes moves and reductions
for the sake of a computation or an evaluation. But this is not always so.
For example, the skein algorithms such as the bracket polynomial algorithm
can be regarded as a reduction process that produces two new graphs from
each crossing in the knot diagram. This is similar to allowing free beta
reduction in the lambda calculus graphs. What must be done however in the
knot theoretic case is to collect up all the end calculation results and
add them together. This is what is meant by a formula like
$$\langle K \rangle = \Sigma_{S} \langle K|S \rangle.$$
(See \cite{kauffman4}.)
Each $S$ is a pattern of calculation leading to a specific algebraic value
$\langle K|S \rangle .$ The topological invariance occurs at the level of the sum of all of
these contributions. 
\smallbreak

\noindent The {\em bracket polynomial} , $\langle K \rangle \, = \, \langle K \rangle (A)$,  assigns to each unoriented link diagram $K$ a 
Laurent polynomial in the variable $A$, such that
   
\begin{enumerate}
\item If $K$ and $K'$ are regularly isotopic diagrams, then  $\langle K \rangle \, = \, \langle K' \rangle$.
  
\item If  $K \amalg O$  denotes the disjoint union of $K$ with an extra unknotted and unlinked 
component $O$ (also called `loop' or `simple closed curve' or `Jordan curve'), then 

$$\langle K \amalg O \rangle \, = \delta \langle K \rangle,$$ 
where  $$\delta = -A^{2} - A^{-2},$$
and
$$\langle  O \rangle \, = 1.$$
  
\item $\langle K \rangle$ satisfies the following formulas 

$$\langle \mbox{\large $\chi$} \rangle \, = A \langle \mbox{\large $\asymp$} \rangle +
 A^{-1} \langle )( \rangle$$
$$\langle \overline{\mbox{\large $\chi$}} \rangle \, = A^{-1} \langle \mbox{\large $\asymp$}\rangle
 + A \langle )( \rangle,$$
\end{enumerate}

\noindent where the small diagrams represent parts of larger diagrams that are identical except  at
the site indicated in the bracket. We take the convention that the letter chi, \mbox{\large $\chi$},
denotes a crossing where {\em the curved line is crossing over the straight
segment}. The barred letter denotes the switch of this crossing, where {\em the curved
line is undercrossing the straight segment}. 
\bigbreak

An analogous situation could occur in GLC  
 where one would need the average over all the results of
the many branching calculations. Note that from a physical point of view,
we are talking about averaging over all the states of a physical system.
Thus in the distributed computing domain, we are interested in finding
ways to collect all the end results, which may well be scattered across
both space and time. Of course, with enough time, all results will be
scattered only across space.
\bigbreak

\begin{figure}[H]
     \begin{center}
     \includegraphics[width=8cm]{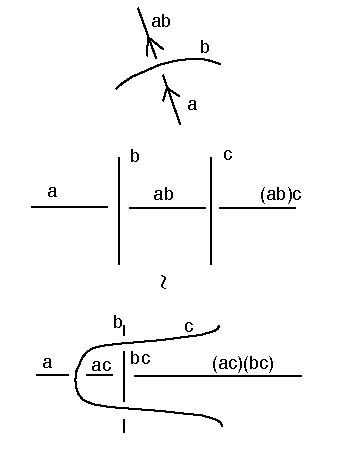}
     \caption{Knot Diagrammatic Multiplication}
     \label{mult}
\end{center}
\end{figure}

We are looking at how one may graft lambda calculus and knot
diagrammatics. Here one can use the knot diagrams at an abstract level to
represent non-associative algebra. See Figure~\ref{mult} for an illustration of the way lines that cross can be used to model multiplication, and for an illustration of how a topological move on the lines corresponds to a self-distributive law: $(ab)c = (ac)(bc).$ 

Knot diagrams can be a space to
write lambda expressions. Then we obtain a graphical lambda calculus that
has the form of knot diagrams equipped with extra lambda nodes and
multiplication nodes, let's call it Topological GLC (TGLC). These diagrams can be transformed by beta moves and
of course the system can be expanded in various ways. So far, to keep the
strictness of the non-associative algebra for lambda calculus, we would
not introduce isotopy moves (Reidemeister moves) on the knot diagrams, but
this can be done if we wish. It leads to a new algebraic investigation
where one is given an algebra with certain axioms (for example rack or
quandle axioms corresponding to Reidemeister moves) and then one
introduces lambda calculus over this algebra. Then, for example,  $G = \lambda x.(ab)x$ is an
operator in the extension of the given algebra that has the property that
$Gc = (ab)c$ for any $c$ in the algebra, and $a,b$ are given elements in the
algebra. We assume that $G$ is now also an element in the algebra.
This means that we assume that $G$ satisfies the relations in the algebra.
This can lead to many algebraic questions difficult to answer. Another way to
proceed is to distinguish lambda operators from the initial algebra and
not demand that they take on the axioms of that algebra. These issues need
to be explored in relation to a topological lambda calculus associated
with knot and link diagrams. We expect that intensive work in the comparison of
quantum link invariants, abstract tensors and lambda calculus will
illuminate many issues related to graphical lambda calculus proper. We are
sure that pursuing this comparison will yield benefits to low
dimensional topology, mathematical foundations  and to the computational
and information processing power of graphical lambda calculus.
\bigbreak

\begin{figure}[H]
     \begin{center}
     \includegraphics[width=8cm]{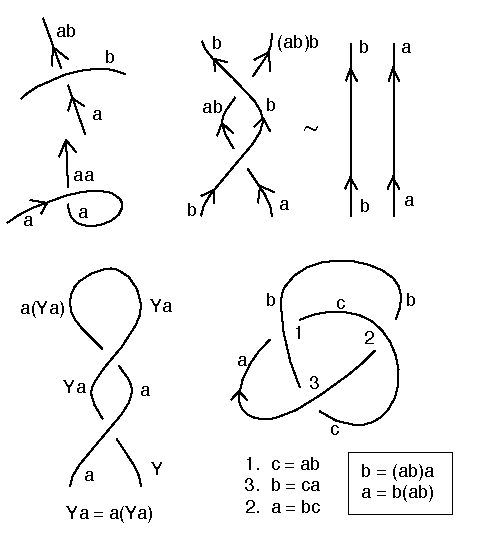}
     \caption{Relations and Diagrams with Loops}
     \label{rel}
\end{center}
\end{figure}

In order to see these issues more clearly, examine Figure~\ref{rel}. In this figure we illustrate how a
simple diagrammatic curl corresponds to the operation of forming the product $aa$ from an element
$a.$ This operation is equivalent to a fan-out combined with a multiplication in the graphical 
methods already discussed in this paper. 

Since self-multiplication is quite important for many lambda calculus constructions, we initially choose to not reduce such curls in the diagrams. This means that we disallow the first Reidemeister move (that uncurls a curl).  The same figure illustrates how when one line passes twice consecutively over another line, then algebraically we get the operation
$(ab)b$ where $b$ is the algebra carried by the overcrossing line and $a$ is the algebra carried by the 
initial undercrossing line. If we wish to have the topological move of separating these lines (as illustrated
in the upper right of the figure), then we need the corresponding algebraic identity $(ab)b = a.$
This identity is not true in a general non-associative algebra. If our algebraic diagrams satisfy both
$(ac)(bc) = (ab)c$ and $(ab)b = b$, then the underlying algebra is called a {\it rack}. We are interested in studying lambda calculus over a rack algebra and we wish to understand if the topological malleability
of these diagrams will lend computing power to the graphical system.

\smallbreak

A further point is illustrated in Figure~\ref{rel}. In the left-bottom of the figure, we see that a double-curl
with labels $a$ and $Y$ at the bottom, leads to the equation $$Ya = a(Ya)$$ by following the identifications indicated in the diagram. Thus we see that certain fixed point expressions are naturally
articulated with the diagrams. In this case $Y$ is the well-known Church-Curry combinator that produces
a fixed point for any $a$ in the lambda calculus. The way such fixed point expressions are produced 
from the knot diagrams (or tangle diagrams where there are free ends) is not dependent upon the Reidemeister moves. These fixed points occur in the free non-associative algebra that labels such diagrams. Another example of this phenomenon is given in the lower right of Figure~\ref{rel}.
Here we show the diagram of the trefoil knot $T$ with labels $a,b,c$ on its arcs. Then corresponding
to the three crossings in the diagram we have the three relations
$$c = ab, b = ca, a = bc.$$ Substituting $c = ab $ into the second two equations, we have
$$b = (ab)a$$ and $$a = b(ab).$$ Thus we see that the trefoil diagram {\it inherently} embodies two
algebraically linked fixed point expressions.

This occurs in the universal associative algebra, before applying any Reidemeister moves to the diagrams or any rack axioms to the algebra. The situation is
analogous to that studied by Aczel in his non-well-founded set theory \cite{Acz}. Aczel uses graphs with cycles to model sets that are members of themselves or members of each other in a circular pattern.
Similarly, we see that knot diagrams embody the properties of fixed point combinators in the lambda calculus. (One can form a non-standard set theory based on knot diagrams. See \cite{kauffman2}.)This fact of diagrammatic life needs to be studied both in relation to GLC and in relation to the topology of the knots. If we allow the topological moves on the knots we can often use knot theory to show that two knots are topologically different or that  given knot diagram is actually knotted. Such verifications will show that the corresponding fixed point combinators are different and/or non-trivial in the graphical lambda calculus with Reidemeister moves allowed. Thus there is a potentially deep interaction between the properties of lambda calculus algebras and the study of topological types of knots and tangles.
\smallbreak

\begin{figure}[H]
     \begin{center}
     \includegraphics[width=8cm]{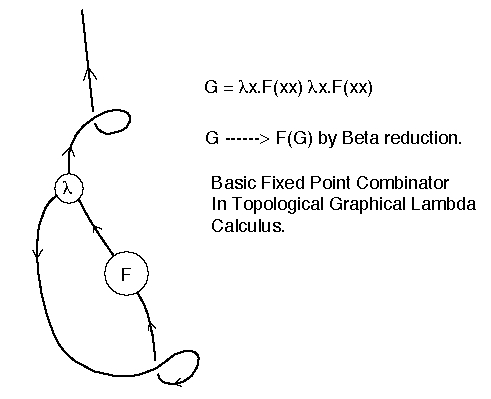}
     \caption{Topological Fixed Point Combinator}
     \label{tfix}
\end{center}
\end{figure}

In Figure~\ref{tfix} we illustrate the basic fixed point combinator 
$$G = \lambda x.F(xx)  \lambda x.F(xx)$$
in (knot diagrammatic) topological graphical lambda calculus ($TGLC$).
the two self-multiplications that occur at two levels in this expression are instantiated by the 
two curls in the graph. We have that $F(G)$ is the beta-reduction of $G$ and thus G corresponds to the fixed point $G = F(G).$ It is important to note that equality in fixed points is translated into
beta-reducibility in the graphical lambda calculus. It is in this way that we can control in a computational system the otherwise infinite loops that could occur if one treated beta reduction as equality. In abstract
algebra the situation is different and one can consider fixed point identities and their consequence for an algebra with generators and relations. The fact that the fixed point combinators can occur both at algebraic and computational levels in the $TGLC$ makes this a rich subject for investigation.
\smallbreak

\begin{figure}[H]
     \begin{center}
     \includegraphics[width=10cm]{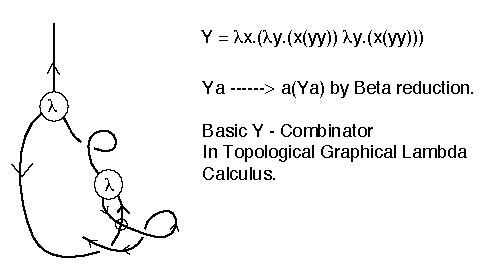}
     \caption{Topological Y - Combinator}
     \label{yfix}
\end{center}
\end{figure}

Similarly, in Figure~\ref{yfix} we illustrate a $TGLC$ expression for the $Y$-combinator.
Note how the structure of this combinator takes on the hybrid nature of tangle diagram infused with curls and lambda nodes. The encircled crossing is a {\it virtual crossing}, a crossing of graphical lines that does not affect them in any way. It is natural to use such a vertex in graph theory and in fact there is an extension of knot theory \cite{kauffman5,kauffman6}  that allows exactly such virtual crossings in the knot diagrams. Thus $TGLC$ can be viewed as a computational extension of virtual knot theory.

Finally, we examine relations of TGLC with topological quantum computing. We point out that a quantum computer
is modeled by a graphical network that embodies the mechanism of unitary
transformations on a complex vector space (Hilbert space). There is a
special reduction move called {\it measurement}  that projects a 
state vector to one of its basis vectors with probability the absolute
square of the coefficient of that basis vector. Such reductions are
usually modeled at the algebraic level. One may look for a more graphical
model for the quantum measurement so that it comes in line with beta
reduction. This involves using tensor networks analogous to the knot diagrammatic networks discussed above, but interpreted in terms of quantum amplitudes. We will look at quantum networks from this
point of view and find ways to formulate quantum computing that interfaces it with graphical lambda calculus.

\end{document}